# Human Assisted Science at Venus: Venus Exploration in the New Human Spaceflight Age

A White Paper for the Planetary Science and Astrobiology Decadal Survey 2023-2032

**Noam. R. Izenberg**  (noam.izenberg@jhuapl.edu, 443-778-7918, Johns Hopkins University Applied Physics Laboratory (JHUAPL), Laurel, Maryland, USA)

**Ralph. L. McNutt Jr.**  *(JHUAPL Laurel, Maryland, USA)*
**Kirby D. Runyon** *(JHUAPL Laurel, Maryland, USA)*
**Paul K. Byrne** *(North Carolina State University, NC, USA)*
**Alexander MacDonald** (NASA Headquarters, Washington, DC, USA)

**Cosigners:**

Jennifer Whitten (Tulane University)
Constantine Tsang (Southwest Research Institute)
Jonathan Sauder (Jet Propulsion Laboratory)
Stephen Kane (Univ. California Riverside)
Dmitry Gorinov (IKI - Space Research Inst., Russia)
Shannon Curry (Univ. California, Berkeley)
Darby Dyar (Planetary Science Institute)
Ye Lu (Kent State University)

Joe O'Rourke (Arizona State Univ.)
Chuanfei Dong (Princeton Univ.)
Ryan McCabe (Hampton Univ.)
Pat Beauchamp (Jet Propulsion Laboratory)
Jeremy Brossier (Wesleyan University)
Thomas Greathouse (Southwest Research Institute)
Kandis-Lea Jessup (Southwest Research Institute)
Scott Guzewich (NASA Goddard Space Flight Ctr.)

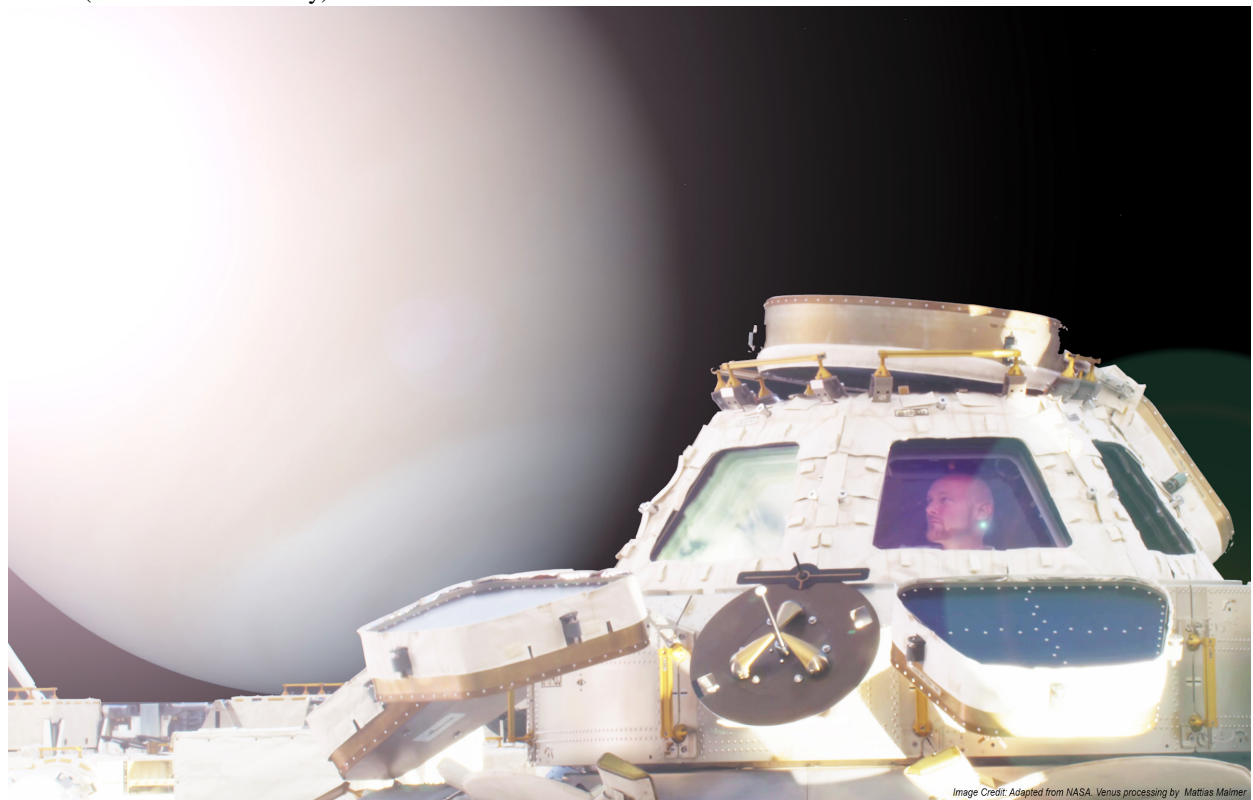

*An imagined scene of an astronaut viewing Venus during a flyby while operating in situ Venus vehicles. Venus image credit: NASA/JHUAPL/MESSENGER/Mattias Malmer. Space Station Image of ESA astronaut Alexander Gerst courtesy NASA.*

**Key Points**
- **Some human mission trajectories to Mars include flybys of Venus**
- **These flybys provide opportunities to practice deep space human operations, and offer numerous safe-return-to-Earth options, before committing to longer and lower-cadence Mars-only flights**
- **Venus flybys, as part of dedicated missions to Mars, also enable "human in the loop" scientific study of the second planet**
- **The time to begin coordinating such Earth-to-Mars-via-Venus missions is *now***

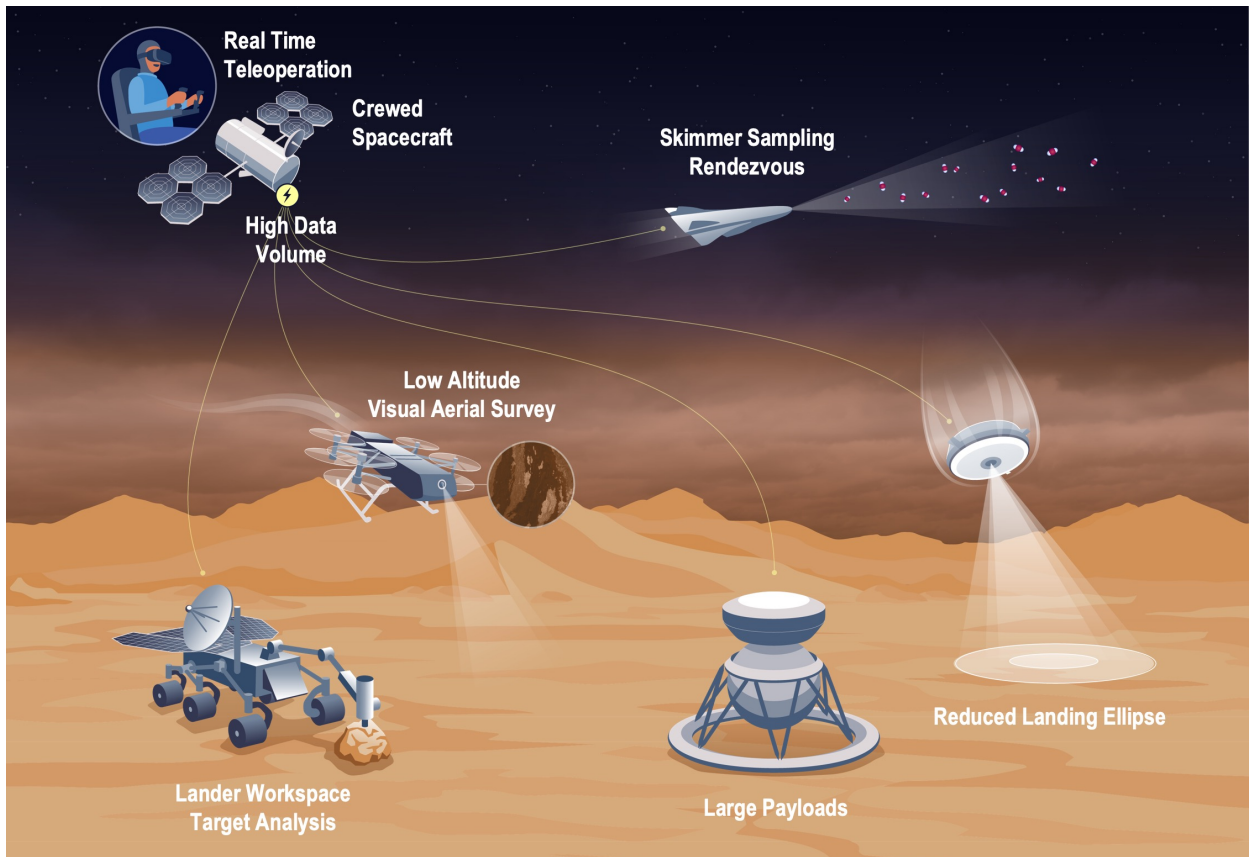

*Figure 1*. *Science enabled and maximized by Human-crewed Venus flyby. Crew module adapted from Cassidy et al., 1967. Image by APL/Caleb Heidel.*

Human Assisted Science at Venus: Venus Exploration in the New Human Spaceflight Age

**Introduction:**

The NASA Administrator's IAC speech in 2019 indicated NASA is considering opposition-class human missions to Mars—where Mars and Earth are close to each other in their orbits at launch —that would include a Venus fly by as part of an overall two-year mission [1]. NASA's recently released "Sustained Lunar Exploration and Development" report [2] confirms that a two-year round-trip mission to Mars, with a short stay on the Martian surface, is the current plan for the first human mission to Mars, implying an opposition-class mission which a Venus flyby can help enable (Figure 2).

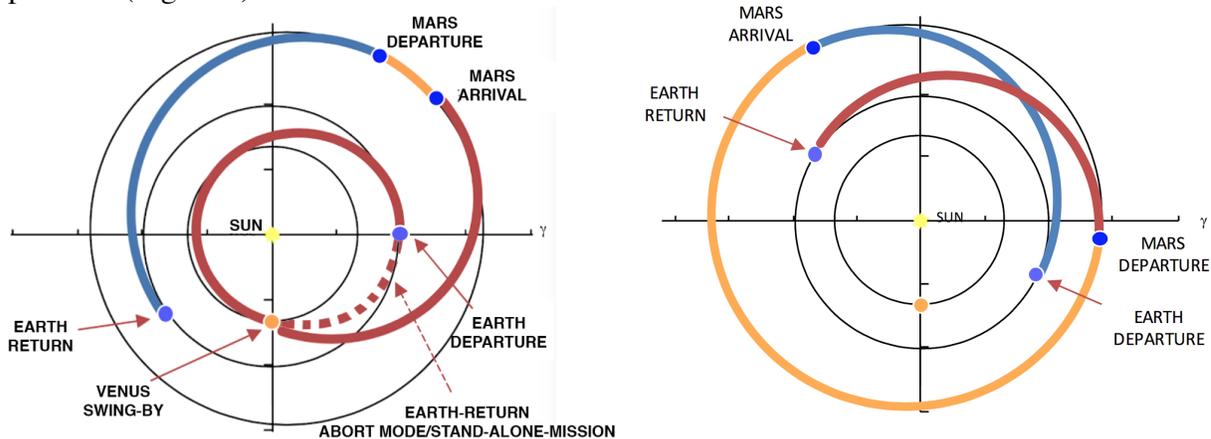

*Figure 2. (Left) Opposition- ("short stay") and conjunction-class ("long stay") Mars missions, respectively, adapted from the Mars Design Reference Architecture (2014) [3]. The opposition mission (left) is Earth-Venus-Mars and features an outbound Venus flyby en route to Mars. Optional Earth-return from Venus as a stand alone Earth-Venus Earth (EVE) mission or Earth-Venus-Mars abort case is shown with dotted line.*

Venus flybys can enhance round-trip Mars exploration. Ever since NASA's EMPIRE ("Early Manned Planetary–Interplanetary Roundtrip Expeditions") and UMPIRE ("Unfavorable Manned Planetary–Interplanetary Roundtrip Expeditions") studies in the 1960s and early 1970s, it has been clear that Venus flybys can reduce the overall energy requirements for opposition-class missions to Mars (Figure 2) [3-9]. **Therefore, a human Venus flyby is not only a 'free' add-on, but a beneficial addition to any opposition-class Mars mission architecture.** As a result, if NASA's first mission to Mars is an opposition-class mission, it is likely that this mission will also include a flyby of Venus. There is reason to be excited by this "two planets for the price of one-plus" approach. A dedicated year-long Venus flyby mission could serve as a valuable "shakedown cruise" for the deep-space transport systems needed for the first human mission to Mars [10]. Further in the future, there is considerable potential for dedicated crewed orbital missions to Venus, including atmospheric exploration [11,12]. All of these opportunities mean that space exploration missions that necessarily include human proximity to Venus should be considered as attractive options for the future of human spaceflight. **We therefore propose that the planetary science community and the decadal planning process take into consideration and support the science planning of efforts that take advantage of human proximity to Venus for expanded scientific discovery.**

**The Science Case for Human Flybys of Venus**

An overarching goal in Venus science is to understand divergent evolutionary paths in the first billion years between Venus, Mars, and Earth [13]. Understanding the modern Venusian atmosphere, surface, and interior are the first steps in piecing together Venus' early history and determining how ancient Venus compared with ancient Earth and Mars. The Venus Exploration Analysis Group (VEXAG) recently listed three overarching goals in Venus science (VEXAG Goals, Objectives and Investigations (GOI) [14]): I) Understand Venus' early evolution and potential habitability to constrain models of the evolution of Venus-sized exo-planets; II) Understand atmospheric composition and dynamics on Venus; and III) Understand the geologic history preserved on the surface of Venus and the present-day couplings between the surface and atmosphere. **Crewed Venus flyby missions enable multiple science mission scenarios not accessible to robotic spacecraft alone, and essentially represent force multipliers in efforts to achieve these major Venus exploration goals.** Having a crewed spacecraft en route to, during, and after a Venus flyby enables new mission architectures (Fig. 1), including, but not limited to:

*Decisions and interaction in real-time.* A piloted flyby of Venus could spend approximately two to over four hours between 8 Venus radii ($R_v$) inbound to 8 $R_v$ outbound at a flyby speeds of approximately 6-14 km/sec, depending on mission design [3, 15-19], with near-zero communications latency. Further, two-way light time for communications would be under one second within ~25 $R_v$ or for approximately six to twelve hours on each of the inbound and outbound portions of the flyby. Very low latency to effective real-time telemetry would allow human-in-the-loop guidance, decision-making, and interactive science for robotic operations in the Venus environment by crew up to days around closest approach. These mission concepts include guideable or flyable aerial platforms [20-22] to surface rovers [23]. Benefits to identified VEXAG GOI science Objectives at Venus include (https://www.lpi.usra.edu/vexag/reports/VEXAG_Venus_GOI_Current.pdf):

o <u>Landing error-ellipse reduction.</u> Final-leg descent/landing components of short- or long-duration landed assets could take advantage of human remote piloting during final surface approach. Altimetric and (when within direct viewing distance of the surface) optical feedback in real time would enable a crew member to guide the final descent as if controlling a remotely-piloted vehicle on Earth, mitigate possible landing hazards, and select optimum landing location. This approach could reduce landing ellipse error and landing hazard in geologically complex terrain including tessera, rift zones, or volcano fields/flanks. (GOI relevance: Location dependent Goals **III.A, III.B;** Venus geologic and atmosphere-surface processes.)

o <u>Low-altitude survey.</u> Using the same "human in the loop" concept, an astronaut flying past Venus could actively conduct an aerial geological imaging/morphology/spectroscopy survey from a low-altitude guideable platform or descent vehicle designed for hours-long, lateral transits during descent. Trained humans are more capable of making informed science choices in a one-time aerial survey than current, autonomous control systems. Depending on the goal(s) and instrument payload, e.g., physical geology (imagers/altimeters), surface magnetic anomaly (magnetometers), gravity survey (accelerometers), composition (imaging spectrometers), real-time scientist-astronauts could maximize science return for multiple Goals. (**I.A, I.B, III.A, III.B;** Venus history of water, habitability, and planetary evolution in context.)

o <u>Rapid in-situ science.</u> High-capability instruments (e.g., LIBS-Raman spectrometers, alpha particle X-ray spectrometers, UV-NIR imagers, etc.) are at present only capable of operating on the Venus surface and near-surface for ~1–2 hours. Some science instruments such as drills or other sample collection devices face a single or small number of opportunities in a limited target area or workspace. Real-time communication between a scientist in the flyby crew and a landed spacecraft on the surface would allow the astronaut to rapidly assess and act upon imaging and compositional data to select targets for maximum science return—and to avoid targeting errors such as unintentional sampling, as occurred to the Soviet's Venera 14 lander [24]. (**I.B, II.B, III.A, III.B**; baseline and variation of processes on, in, and around Venus.)

*Sample Retrieval*. Atmospheric skimmers have been proposed for active sampling of the upper atmosphere of Venus and its near-space environment [25, 26]. Such fast sample-grab-and-return from the Venus atmosphere, rendezvousing with the departing spacecraft instead of transiting to Earth, may retire at least some risk such missions face. If atmospheric or other kinds of sample retrieval is possible during a piloted flyby, the involvement of a human element to ensure rendezvous and capture with a sampling device/mechanism/sub-satellite could improve chances of sample return, and potentially allow the astronaut themselves to conduct science research on those samples during their journey. (**I.A., I.B, II.B,** in GOI).

*Mission Infrastructure*. Considerable telecommunications and data storage capability will be required for any interplanetary human mission, and thus would be available opportunistically during a Venus flyby en route to Mars, or intrinsically for an EVE mission. These assets would reduce the distance that signals need to travel from the Venusian atmosphere or surface to a spacecraft and potentially reduce systems costs or enable new classes of payload. Indeed, such infrastructure might potentially increase the data volume that can be acquired from in-situ instruments at Venus.

*Spacecraft Mass*. A crewed mission to beyond near-Earth space, much less Mars, is no small affair. Between fuel, habitation, protection, power, communication, and human-rated safety requirements, any deep Solar-System-capable mission will require sending many tonnes of material on interplanetary trajectories. Although "every kilogram counts" on such missions, when the baseline mission is massive, the mass cost of marginal science packages and probes for Venus is comparatively small.

*Assembly en route*. Not specific to any particular science goal, but potentially enabling delivery of larger and more complex science capabilities to Venus, individual components of large, modular probes or orbiters may be launched from Earth separately and assembled by crew en route to Venus without the need for fine remote manipulation or machine intelligence. Mission concepts include CubeSat – or larger – constellations [*e.g.* 27].

**The Human Case for Human Flybys of Venus**

Venus flybys directly enable shorter, less expensive crewed missions to Mars—and, as preparation, humanity's first planetary mission beyond the Earth–Moon system could feature a one-year flyby exploration mission of Venus: an "Apollo 8 analog" for deep-space human flight past a planetary target. **Such an Earth–Venus–Earth (EVE) mission also practices the only return-to-Earth abort mode a Mars-bound crew would have.** Not only would such a mission provide exploration and science opportunities at and on Venus, but would also serve as valuable deep space practice for the first humans-to-Mars mission, on a more rapid cadence than an Earth–Mars flyby can be accomplished. If the current two-year baseline mission to Mars [28] is adopted, it will be an opposition mission and thus a Venus flyby is automatically an enabling option, depending on the launch window. This requirement might in fact make an EVE "shakedown" mission even more valuable than a conjunction-class Mars mission for reducing technical risk and gaining interplanetary flight experience.

*Communications Latency:* On or orbiting Mars, astronauts will experience telecommunications latency with Earth given the finite speed of light. The round-trip light travel time between Earth and Mars ranges from 6.4 minutes to 44 minutes, making real-time conversations impossible. For Venus, the same metric ranges from 4.6 minutes to 28 minutes [29]: comparable to, yet lower than, the Mars case. A human Venus mission will thus afford us practice in dealing with this communications latency—and at the same time providing a compelling exploration destination other than empty space (such as a Lagrange point). In analog simulations, spacewalking astronauts have found that the equivalent of text messaging is an effective way to communicate when dealing with such latency [31].

*Crew Health:* There is at present little information on the physiological and cognitive effects of long-duration spaceflight outside Earth's magnetosphere on mission scenarios without a rapid Earth-return capability. Although a human Venus flyby mission would be shorter than even a human Mars flyby mission, there would nonetheless be no opportunity to return to Earth quickly, unlike Low Earth Orbit and lunar missions. Compounded by the telecommunications latency, such a mission would provide considerable psychological and emotional stressors, but at the same time the opportunity to characterize and mitigate those stressors for subsequent, longer-duration Mars missions. Long-duration isolation studies show how combinations of routine, leisure time, and meaningful work [31-34] contribute to maintaining crew mental and physical health. A science-oriented Venus flyby would provide not only a focus of meaningful work around the halfway point of an EVE flyby (or the outbound leg of an Earth–Venus–Mars–Earth mission), but would create a unique meaningful work scenario—work that only the crew of that mission could accomplish during the planetary flyby.

*Accessibility and Radiation:* Venus, on average, is much closer to Earth (1.12 AU) than Mars (1.69 AU), allows for shorter overall mission durations (thus simplifying crew logistics and time in space), and has more frequent planetary alignments than Mars (every 19 months versus 26 months for Mars). A human Venus flyby mission would take less than a year—shorter than some missions to the International Space Station—yet would expose the crew to higher solar radiation levels comparable to those on a flight to Mars, albeit slightly reduced galactic cosmic radiation due to solar shielding [35-37]. An EVE flyby mission would expose astronauts to a similar total dose of radiation as an Earth-Mars-Earth (EME) flyby-only mission, and even the longer EVME opposition Mars missions of ~700 days would result in less total exposure than the shortest EME conjunction missions of 850 days [36].

*The Practice Effect:* Science activities and communications during a human flyby of Venus will in many ways be a dress rehearsal for Mars. In the case of an EVME mission, some operations during approach and flyby Venus will be very similar to those at Mars (with the notable exception of orbit insertion for an orbital/landed Mars mission). In a possible EVME dual-flyby precursor, or if the orbital phase of a Mars rendezvous mission were aborted for any reason, the Mars flyby would be operationally similar to a Venus flyby, possibly including delivery and teleoperation of robotic probes.

**The Programmatic Case for Human Flybys of Venus**

Getting humans to Mars is a fundamental, mandated goal of NASA. **Going to Venus helps us get to Mars every bit as much as going to the Moon does within the "Moon to Mars" paradigm.**

*Interdivision interdependence*: The opportunities for crewed missions to Venus extend beyond the compelling scientific investigations such missions would offer. To ensure the safe travel of humans near Venus, considerable cross-divisional collaboration within NASA is required. NASA's Human Exploration and Operations Mission Directorate (HEOMD) will likely lead these efforts, but the Science Mission Directorate (SMD) and Space Technology Mission Directorate (STMD) have key roles to play. **Working to coordinate all three Mission Directorates in support of flying humans to Venus, either as a dedicated mission or as part of a flight to Mars, will set a strong basis for future, continued collaboration and perhaps even for future public–private partnerships engaged in the exploration of space.**

For example, whereas HEOMD will have a science focus on human physiology, psychology, and health over long-duration deep space missions, SMD can offer to define planetary and heliophysics science investigations. Similarly, SMD has substantial experience operating in deep space and particularly in the Venus environment, with a legacy of missions from Mariner 2 through Parker Solar Probe. Providing engineering insight into thermal design, telecommunications, and navigation, in addition to developing multidisciplinary research programs for both cruise and Venus flyby phases, will require that SMD closely work with HEOMD from the outset of any serious discussion of sending humans to (or past) Venus. Moreover, where there are issues that extend beyond the purview of both HEOMD and SMD, STMD will be able to contribute expertise, e.g., regarding the development of new or enhanced technologies to support safe human operations tens of millions of km from Earth. Since such technologies may also be required to support or at least accommodate scientific measurements (e.g., via the placement of instruments on the spacecraft exterior), effective coordination between SMD and STMD will be crucial. Close coordination between NASA's HEOMD, SMD, and STMD will be key to successfully achieving human exploration of the Venus *and* Mars environments. The sooner such coordination can begin, no matter how far away such prospective missions are, the better.

*Economics of two-planet missions:* A science-intensive flyby of Venus on any leg of a crewed mission to Mars makes any such endeavor a two-planet mission. Venus science is every bit as interesting and important as Mars, and just as crucial to comparative planetology. A detailed cost analysis of this science scenario is needed, but it is likely that a 'two-planet mission' approach would more than justify adding a Venus science focus to Mars-bound flybys of Venus.

**FINDING: "Humans to Mars—Via Venus" is logical, exciting, and offers unprecedented science at Mars and Venus at a fraction of the cost of dedicated crewed missions to both.**

***Venus is how we get to Mars.***